\documentclass[aps]{revtex4}
\usepackage{graphicx}
\usepackage{bm}
\begin{document}
\newcommand{\area}{{A}}
\newcommand{\deltaK}{{K'}}
\newcommand{\DeltaV}{\Delta V}
\newcommand{\Is}{{I_{\rm s}}}
\newcommand{\jc}{{j^{\rm cr}}}
\newcommand{\js}{{j_{\rm s}}}
\newcommand{\jsc}{{j_{\rm s}^{\rm cr}}}
\newcommand{\Kp}{{K_\perp}}
\newcommand{\kB}{{k_B}}
\newcommand{\kv}{{\bm k}}
\newcommand{\Mw}{{M_w}}
\newcommand{\Mphi}{{M_{\phi}}}
\newcommand{\phiz}{{\phi}}
\newcommand{\Sv}{\bm S}
\newcommand{\sigmav}{{\bm \sigma}}
\newcommand{\vc}{{v_{\rm c}}}
\newcommand{\ve}{{v_{\rm e}}}
\newcommand{\Vz}{{V_0}}
\newcommand{\Vpin}{{V_{\rm pin}}}
\newcommand{\Vphi}{{V_{\phi}}}

\title{ 
Universality of thermally assisted magnetic domain wall motion
under spin torque 
}
\author{Gen Tatara}
\altaffiliation[
Work done at the Laboratoire de Physique des Solides, Universit\'e Paris-Sud, 
Orsay, France
]{}
\affiliation{
PRESTO, JST, 4-1-8 Hooncho Kawaguchi, Saitama, Japan \\
and\\
Graduate School of Science, Osaka University, Toyonaka, Osaka 560-0043, 
Japan}
\author{Nicolas Vernier and Jacques Ferr\'e}
\affiliation{
Laboratoire de Physique des Solides, UMR CNRS 8502, Bat 510, University Paris-Sud, 91405, 
Orsay, France
}

\date{\today}

\begin{abstract}
Thermally assisted motion of magnetic domain wall under spin torque is studied
 theoretically. 
It is shown that the wall velocity $v$ depends exponentially on the spin current,
 $\Is$, below the threshold value, in the same way as in a 
 thermally activated motion driven by a force.
A novel property of the spin torque driven case at low temperature is that the linear term in 
 spin current is universal, i.e., 
$\ln v \sim \frac{\pi\hbar}{2e}(\Is/\kB T)$.
This behavior, which is independent of pinning and material constants, could
 be used to confirm experimentally the spin torque as the driving
 mechanism. 
\end{abstract}
\maketitle

Electric control of domain wall dynamics in nano-ferromagnets is currently 
under intensive study from the viewpoint of both applications and
fundamental physics
\cite{{Grollier02,Tsoi03,Klaeui03,Vernier04,Yamaguchi04,Yamanouchi04,Lim04,Saitoh04},
{TK04,Thiaville04,ZhangLi04}}.
In most ferromagnetic wires so far studied, domain wall motion  
driven by a steady current is  probably related to 
 the spin torque induced mechanism by a spin current, since domain walls are thick
 compared with Fermi wavelength, and thus the momentum transfer effect
would be neglected\cite{TK04}.
Experimentally, the critical current needed to drive the wall is of
most interest at present, but the wall velocity would be essential
to state the involved mechanism, and also to imagine new applications.
So far, the wall velocity has not been much investigated experimentally. 
Yamaguchi et al. performed measurements on a narrow permaloy
wire\cite{Yamaguchi04}, 
but the data were not precise enough to determine unanbiguously the dependence on the applied current.
Clearer results were obtained recently on magnetic semiconductor wires 
\cite{Yamanouchi04,Yamanouchi04a}, where an exponential dependence of the wall
velocity as function of the applied current density ($j$) was evidenced to be $\ln v \propto  j$
\cite{Yamanouchi04a}.
This behavior is reminiscent of a thermally assisted motion under a driving force, 
often found in many systems below its threshold value\cite{Bruno90,Ferre02,Muller01}.

An interesting question here is whether the thermally activated motion 
under spin torque is indeed similar as that induced by a force. 
The answer is not obvious since domain wall dynamics 
induced by a spin torque above the threshold current 
has shown to behave quite differently from a force driven case\cite{TK04}.
In the case of a uniform magnetization resersal, 
Li and Zhang\cite{Li04} found theoretically that  spin dynamics under a spin torque
are characterized by an exponential dependence of the flip rate 
on the spin torque, i.e., a similar behavior as in a field-driven case.

\begin{figure}[bp]
\includegraphics[scale=0.4]{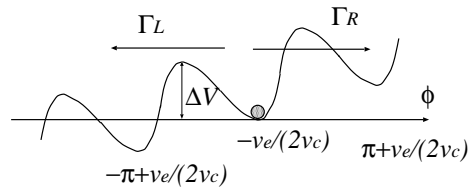}
\caption{Potential for $\phiz$ under spin current.
\label{FIGphipot}}
\end{figure}

Let us consider a planar domain wall at position $X(t)$ moving in one
direction in the presence of a pinning potential of harmonic type
\begin{equation}
\Vpin(X)=\frac{1}{2}\Mw \Omega^2 X^2 \theta(\lambda-|X|),
\label{vpin}
\end{equation}
where $\Mw$ is the wall mass,
$\lambda$ the wall thickness,  
$\Omega$ the pinning frequency, and $\theta(x)$ the step function.
The mass of the wall is related to the hard axis anisotropy, $\Kp$, as 
$\Mw= \frac{\hbar^2 N}{\Kp \lambda^2}$\cite{Chikazumi97} where
$N\equiv 2\lambda \area/a^3$ is the number of spins in the wall.
The coupling to the electric current is induced by the exchange
interaction.  For a thick wall,  
which is usually the case in most experiments, the domain wall is driven
by a spin torque effect\cite{TK04}.
Introducing $\phiz$ as average of the spin angle from the easy plane, the equations of motion that describes wall dynamics can be written as\cite{TK04}
\begin{eqnarray}
\dot{\phiz}+\alpha\frac{\dot{X}}{\lambda}
 &=& -\frac{1}{2}\frac{\Omega^2}{\vc} X  \nonumber\\
\dot{X}-\alpha\lambda\dot{\phiz} 
 &=& \vc\sin2\phiz+\ve,
\label{eqxphi}
\end{eqnarray}
where $\vc\equiv \Kp\lambda S/(2\hbar)$, 
$\ve\equiv \frac{a^3}{2Se}\js$ is the drift velocity of the spin current
density ($\js$), which describes the spin torque effect.
$\alpha$ is the Gilbert damping parameter.
Dynamics of the wall under a spin torque is calculated in
terms of the dynamics of $\phiz$, as pointed out in Ref. \cite{TK04}.
From the above 
set of equations we obtain a closed equation of motion for $\phiz$
as
\begin{equation}
(1+\alpha^2)\Mphi \ddot{\phiz}
  = -\alpha\Mphi \dot{\phiz}\left(
  \frac{\lambda\Omega^2}{2\vc}+\frac{2\vc}{\lambda}\cos2\phiz \right)
  -\frac{NS^2\Kp}{2}\left(\sin2\phiz +\frac{\ve}{\vc}\right),
\label{eqofmo}
\end{equation}
where the effective mass of $\phi$ is given by\cite{TT96}: 
$\Mphi\equiv NS^2\Kp/\Omega^2$.
We see that 
the potential energy for $\phiz$ is given by
a tilted washboard potential,
$\Vphi=\frac{NS^2\Kp}{2}\left(\sin^2\phiz +\frac{\ve}{\vc}\phiz\right)$,
and the oscillation frequency of $\phiz$ around its minimum
is given by
$\Omega/\sqrt{1+\alpha^2}$  for $\ve\simeq 0$ (see Fig. \ref{FIGphipot}).
We consider the case of a weak spin current (much smaller than the
threshold value (i.e., $\js \ll \jsc$, 
$\jsc\equiv\frac{eS^2}{\hbar a^3}\Kp\lambda$\cite{TK04}), which 
allows to linearize the potential with respect to $\js$ (or $\ve$).
The minimum of the potential is then given by $\phiz\simeq -\frac{\ve}{2\vc}$,
and its maxmum occurs for $\phiz\simeq -\frac{\pi}{2}+\frac{\ve}{2\vc}$.
For $\phiz$, the barrier height, $\DeltaV$,  thus reads
\begin{equation}
 \DeltaV= \frac{NS^2\Kp}{2}\left(1-\frac{\pi}{2}\frac{\ve}{\vc}
  + O(\frac{\ve}{\vc})^2 \right).
  \label{linealize}
\end{equation}
Note that this barrier is not due to the pinning ($\Omega$), but
 is related to the anisotropy energy.  
This is an important property of spin torque induced wall motion.
Here we neglect the damping term in Eq. (\ref{eqofmo}), considering
a fast activation process.
This is true if the activation process occurs in a time shorter than 
$\tau=\frac{2\vc}{\lambda\Omega^2 \alpha}$ (for more details, see below). 
The transition rate from $\phiz=-\frac{\ve}{2\vc}$ to the next minimum 
$\phiz\sim -\pi+\frac{\ve}{2\vc}$, induced by thermal activation 
is thus given by
\begin{equation}
\Gamma_L=\Omega e^{-\DeltaV/(\kB T)}
 =\Omega e^{-NS^2\Kp/(2\kB T)} e^{\frac{\pi}{2} \hbar \Is/(e \kB T)},
 \label{v}
\end{equation}
where $\Is\equiv \area \js$ is the spin current.
The hopping probability in the other direction (i.e., to the minimum at
$\phiz\sim \pi+\frac{\ve}{2\vc}$)
is given by
\begin{equation}
\Gamma_R=\Omega e^{-NS^2\Kp/(2\kB T)} e^{-\frac{\pi}{2} \hbar \Is/(e \kB T)},
\end{equation}
and thus,
the average drift velocity of $\phi$ is given by the difference,
$\Gamma\equiv (\Gamma_L -\Gamma_R)$, as
$\langle \dot{\phiz} \rangle \simeq -\pi \Gamma$. 
This last velocity is proportional to the domain wall velocity \cite{TK04} 
(except in the extremely strong pinning case, where $\Vz > \Kp/\alpha$, if $\Vz$
stands for the depth of the pinning potential).
By use of $\dot{\phiz}=-\alpha\frac{\dot{X}}{\lambda}$, after
depinning (eq. (\ref{eqxphi}) with $\Omega=0$), 
the wall velocity is given by  
\begin{equation}
v\equiv \langle \dot{X} \rangle \simeq \frac{\pi\lambda}{\alpha}\Gamma
=\frac{2\pi\Omega\lambda}{\alpha} e^{-NS^2\Kp/(2\kB T)} 
 \sinh \frac{\pi}{2} \frac{\hbar \Is}{e \kB T}.
\end{equation}
It is notable that Logarithm of the velocity shows an universal dependence on
$\Is$ (i.e., it does not involve material parameters) 
\begin{equation}
\ln v \simeq
 \ln \sinh \frac{\pi}{2} \frac{\hbar \Is}{e \kB T}
+C(T),
\end{equation}
where $C(T)=-\frac{NS^2\Kp}{2\kB T}+\ln \frac{2\pi\Omega\lambda}{\alpha}$ 
is independent of $\Is$.
At low enough temperature, $\frac{\hbar \Is}{e \kB T} \gg 1$,
a typical exponential dependence on spin torque is evidenced
\begin{equation}
v =e^{C(T)/2}  e^{\frac{\pi}{2} \frac{\hbar \Is}{e \kB T}},
\end{equation}
which is similar to the thermally activated
wall motion under a small force.
The dependence of $\ln v$ on $\Is$ then writes  
\begin{equation}
\ln v \simeq
\frac{\pi}{2} \frac{\hbar \Is}{e \kB T}
+C'(T), \label{lnv}
\end{equation}
with a coefficient $\frac{\pi\hbar}{2e}$ independent on 
$\Omega$, $\Kp$, nor on $\lambda$  ($C'(T)\equiv C(T)-\ln 2$). 
This universal behavior in the thermally activated regime would be used to identify the driving
mechanism (i.e., spin torque or momentum transfer). 

In deriving eq. (\ref{v}), we neglected the dissipation, that is
justified if the disspation time scale, 
$\tau=\frac{2\vc}{\lambda\Omega^2\alpha}$ is longer than the activation
time, $\Gamma^{-1}=\frac{\pi\lambda}{\alpha v}$. 
Thus, the above statement only applies if the domain wall velocity satisfies 
\begin{equation}
v \gtrsim \frac{\pi}{2}\frac{(\lambda\Omega)^2}{\vc}.
\label{vcondition}
\end{equation}
This condition is easily satisfied. 
For instance, for $\Omega\sim 10$MHz \cite{Saitoh04} and
$\lambda=100$nm, $\lambda\Omega=1$[m/s], while $\vc\sim 600$[m/s] if
$\Kp=0.1$K per site, that gives 
$\frac{\pi}{2}\frac{(\lambda\Omega)^2}{\vc}\sim 3\times 10^{-3}$[m/s], while 
experiments were performed for $v\gtrsim 1$[m/s]\cite{Yamaguchi04,Yamanouchi04}. 

The linearization with respect to $\js$ in eq. (\ref{linealize}) requires 
$\js \ll \jsc$, and this condition and eq.  (\ref{vcondition}) detemine the window where the universal behavior is satisfied. 
This window might be wide as indicated by recent experimental observation of 
the exponential dependence of $v$ on $\Is$  
in GaMnAs\cite{Yamanouchi04a} for a current density larger than 
$10^9$[A/m$^2$] at tempratures in the $86-94$K range.

In conclusion, we have found an universal behavior for the wall  
velocity in the thermally activated regime below the threshold spin current. 
The analysis is based on the standard theory of thermal activation,
which assumes small oscillations in potential wells.
The pinning plays only an essential role in introducing a finite attempt frequency, $\Omega$.


\vspace{5mm}

Acknowledgements: 

The authors are grateful to M. Yamanouchi, D. Chiba, F. Matsukura,
H. Ohno, R. Stamps, H. Kohno and J. Shibata
for valuable discussion.
GT thanks Monka-shou
 and The Mitsubishi Foundation for financial support, 
 and the CNRS for granting his stay in France as an invited associated researcher.



\begin{thebibliography}{99}
\bibitem{Grollier02}
J. Grollier, D. Lacour, V. Cros, A. Hamzic, A. Vaur\'es, A. Fert, D. Adam 
and G. Faini,
J. Appl. Phys. {\bf 92}, 4825 (2002);
Appl. Phys. Lett. {\bf 83}, 509 (2003).

\bibitem{Tsoi03}
M. Tsoi, R. E. Fontana, and S. S. P. Parkin 
Appl. Phys. Lett. {\bf 83}, 2617 (2003). 

\bibitem{Klaeui03}
M. Kl\"aui, C. A. F. Vaz, J. A. C. Bland, W. Wernsdorfer, G. Faini, E. Cambril 
and L. J. Heyderman,
Appl. Phys. Lett. {\bf 83}, 105 (2003).

\bibitem{Vernier04}
N. Vernier, D. A. Allwood, D. Atkinson, M. D. Cooke and R. P. Cowburn, 
Europhys. Lett. {\bf 65}, 526 (2004).

\bibitem{Yamaguchi04}
A. Yamaguchi, T. Ono, S. Nasu, K. Miyake, K. Mibu and T. Shinjo,
Phys. Rev. Lett. {\bf 92} 077205 (2004).

\bibitem{Yamanouchi04}
M. Yamanouchi, D. Chiba, F. Matsukura, and H. Ohno, 
Nature, {\bf 428}, 539 (2004). 

\bibitem{Lim04}
C. K. Lim, T. Devolder, C. Chappert, J. Grollier, V. Cros, A. Vaur\'s, 
A. Fert, and G. Faini, 
Appl. Phys. Lett. {\bf 84}, 2820 (2004).  

\bibitem{Saitoh04}
E. Saitoh, H. Miyajima, T. Yamaoka and G. Tatara, 
preprint.


\bibitem{TK04}
G. Tatara and H. Kohno, Phys. Rev. Lett. {\bf 92} 086601(2004).

\bibitem{Thiaville04}
A. Thiaville, Y. Nakatani, J. Miltat, and N. Vernier,
J. Appl. Phys. {\bf 95}, 7049 (2004).

\bibitem{ZhangLi04}
Z. Li and S. Zhang,
Phys. Rev. Lett. {\bf 92}, 207203 (2004);
S. Zhang and Z. Li,
Phys. Rev. Lett. {\bf 93}, 127204 (2004).


\bibitem{Yamanouchi04a}
M. Yamanouchi, D. Chiba, F. Matsukura and H. Ohno,
private communication.

\bibitem{Bruno90}
P. Bruno, G. Bayreuther, P. Beauvillain, C. Chappert, G. Lugert,
	D. Renard, J. P. Renard and J. Seiden, 
J. Appl. Phys. {\bf 68}, 5759 (1990).

\bibitem{Ferre02}
J. Ferr\'e, in
{\it Spin Dynamics in Confined Magnetic Structures I},
Ed. by B. Hillebrands, Springer, 
Topics in Appl. Phys. {\bf 83}, 127 (2002).

\bibitem{Muller01}
M. Muller, D. A. Gorokhov and G. Blatter,
Phys. Rev. {\bf B63}, 184305 (2001).

\bibitem{Li04}
Z. Li and S. Zhang 
Phys. Rev. {\bf B69}, 134416 (2004).

\bibitem{Chikazumi97}
S. Chikazumi,
{\it Physics of Ferromagnetism} 
(Oxford university press, 1997).


\bibitem{TT96}
S. Takagi and G. Tatara, Phys. Rev. B{\bf 54}, 9920 (1996). 

\end{thebibliography}
\end{document}